\documentclass[review]{elsarticle}

\makeatletter
\def\ps@pprintTitle{%
 \let\@oddhead\@empty
 \let\@evenhead\@empty
 \def\@oddfoot{\centerline{\thepage}}%
 \let\@evenfoot\@oddfoot}
\makeatother

\usepackage{lineno}

\modulolinenumbers[5]
\usepackage{color}
\usepackage{amsmath}
\usepackage{amssymb}

\journal{JOURNAL NAME}

\biboptions{authoryear}

\begin{document}

\begin{frontmatter}

\title{CFD simulation of the wind field over a terrain with sand fences: Critical spacing for the protected soil area}

\author[Fortaleza,Cologne_Geosciences]{Izael A. Lima}
\author[Cologne_Geosciences]{Eric J. R. Parteli}
\author[Cologne_Geosciences]{Yaping Shao}
\author[Fortaleza]{Jos\'e S. Andrade Jr.}
\author[Fortaleza,PMMH]{Hans J. Herrmann}
\author[Fortaleza]{Asc\^anio D. Ara\'ujo\corref{mycorrespondingauthor}}\ead{ascanio@fisica.ufc.br}\cortext[mycorrespondingauthor]{Corresponding author}



\address[Fortaleza]{Departamento de F\'{\i}sica, Universidade Federal do Cear\'a, Fortaleza, 60451-970 Fortaleza, Cear\'a, Brazil}
\address[Cologne_Geosciences]{Department of Geosciences, University of Cologne, Pohligstr.~3, 50969 Cologne, Germany}
\address[PMMH]{PMMH. ESPCI, 7 quai St. Bernard, 75005 Paris, France}

\begin{abstract}
Sand fences are often erected to reduce wind speed, {{prevent aeolian soil erosion}}, and induce sand deposition and dune formation in areas affected by sand encroachment and desertification. However, the search for the most efficient array of fences by means of field experiments alone poses a challenging task given that field experiments are {affected by} weather conditions. Here we apply Computational Fluid Dynamic simulations to investigate the three-dimensional {structure of the turbulent wind field} over an array of fences of different sizes, porosity and spacing. {{Our goal is to perform a quantitative analysis of this structure in the absence of saltation or suspension transport, as first step toward the development of a continuum simulation of the aeolian soil in presence of the fences.}} We find that the area of soil protected against direct {aerodynamic} entrainment has {two regimes, depending on} the spacing $L_x$ between the fences. When $L_x$ is smaller than a {critical value $L_{xc}$}, the wake zones associated with each fence are inter-connected (regime A), while these wake zones appear separated from each other (regime B) when $L_x$ exceeds this critical value of spacing. The system undergoes a second order phase transition at $L_x = L_{xc}$, with the cross-wind width of the protected zone scaling with $\left[{1-{L}_x}/{L}_{xc}\right]^\beta$ in regime A, with $\beta \approx 0.32$. Our findings have implication for a better understanding of aeolian transport in the presence of sand fences, as well as to develop optimization strategies for measures to protect soils from wind erosion.
\end{abstract}


\end{frontmatter}


\section{Introduction}

{Wind-blown sand is one important factor for the erosion of soils, the abrasion of rocks, the morphodynamics of ripples and dunes and the propagation of desertification.} The most important transport mode of wind-blown sand grains is saltation, which consists of grains moving in nearly ballistic trajectories thereby ejecting new grains upon collision with the soil (splash) \citep{Bagnold_1941,Shao_and_Li_1999,Almeida_et_al_2006,Almeida_et_al_2008,Shao_2008,Carneiro_et_al_2011,Carneiro_et_al_2013}. {The splash is also one of the main factors for the emission of dust \citep{Shao_et_al_1993,Shao_2001,Lu_and_Shao_1999}, which,} once entrained, may be transported over thousands of kilometers in suspension, thereby affecting climate and human health \citep{Kok_et_al_2012}. Soil protection from aeolian erosion constitutes, thus, one aspect of broad implication for climate, environment and society. 

To achieve aeolian soil protection, sand fences of various types are constructed with the aim at reducing wind velocity and inducing sand accumulation and dune formation \citep{Li_and_Sherman_2015,Gillies_et_al_2017}. Such fences typically consist of wire, lightweight wood strips or perforated plastic sheets attached to regularly spaced stakes \citep{Pye_and_Tsoar_1990}. Moreover, most sand fences are porous, as solid fences that completely block the wind {may} induce strong vortices extending up to several fence heights downwind, while a porous fence protects larger areas of leeward sheltered ground than its solid counterpart does \citep{Cornelis_and_Gabriels_2005,Bruno_et_al_2018}. 

\begin{figure}[htpb]
\centering
\includegraphics[width=0.6\linewidth]{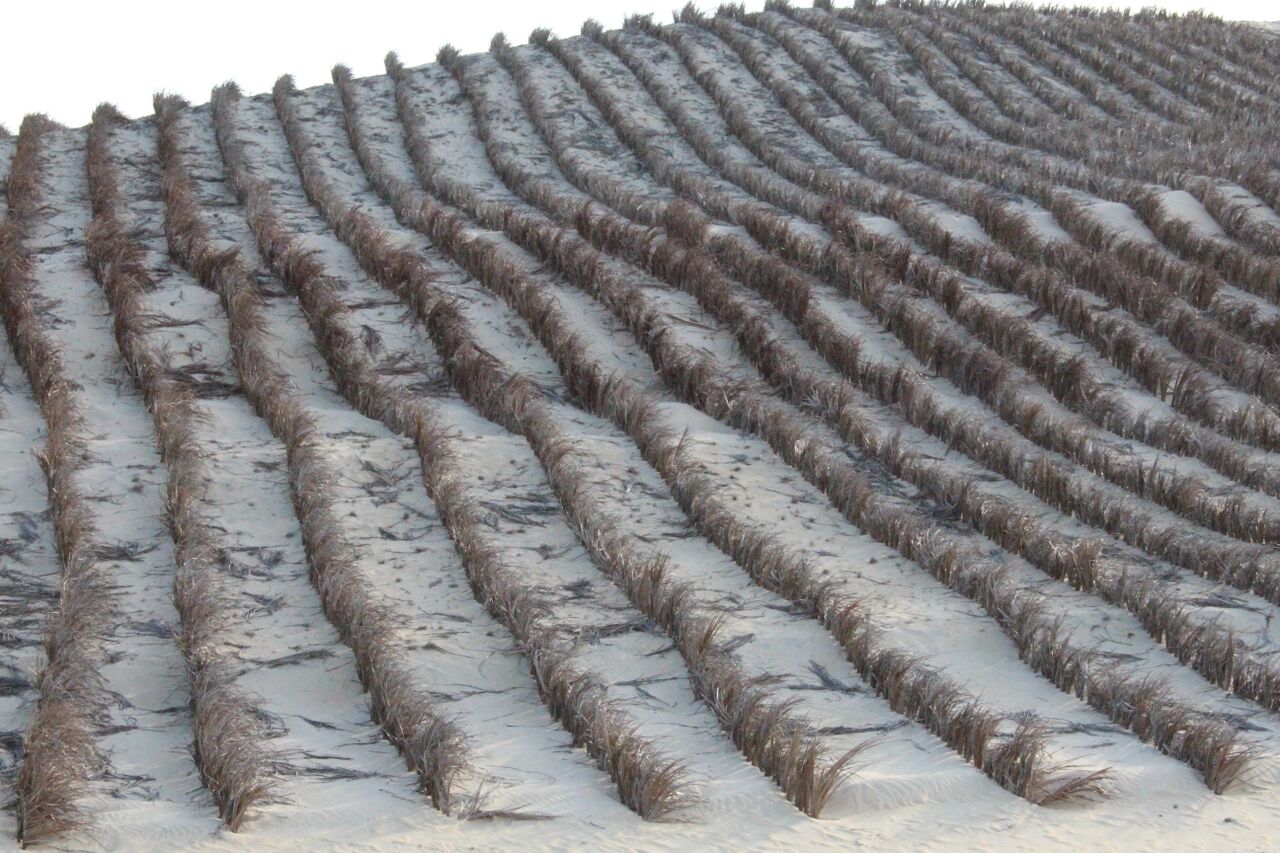}
\caption{Application of sand fences to prevent wind erosion --- a field example. The image shows fences made of coconut leaves in Paracuru, near Fortaleza, main city of State of Cear\'a in Northeastern Brazil (photo by first authors, I.A.L. and A.D.A.).}
\label{fig:sand_fences}
\end{figure}

{{The objective the sand fence array is to reduce the wind velocity below the minimal value required for saltation transport, which is the main sediment transport mode causing aeolian soil erosion. In other words, soil erosion is caused at those places where the wind velocity is above this minimal threshold, which is of the order of 6~m/s (measured at a height of 1~m above the soil) but can be higher depending, for instance, on the presence of moisture or non-erodible elements on the soil \citep{Pye_and_Tsoar_1990,Tsoar_2001,Kok_et_al_2012}. Moreover, soil erosion is particularly strong at coastal and desert areas subjected to unidirectional sand-moving winds. While multidirectional wind regimes lead to accumulating dunes, unidirectional wind regimes cause the formation of migrating dunes \citep{Wasson_and_Hyde_1983}, thus contributing to the propagation of desertification.}}

However, the properties of three-dimensional turbulent wind flow over an array of fences are still poorly described. Their understanding is important to accurately predict wind erosion patterns in the presence of sand fences and to develop efficient strategies to protect sediment soils from aeolian erosion. While most of the previous investigations \citep{Baltaxe_1967,Wilson_1987,Lee_and_Kim_1999,Lee_et_al_2002,Xiaoxu_Wu_2013,Dong_2006,Marijo_Telenta_2014,Ning_Zhang_2015,Tsukahara_et_al_2012,Savage_1963,Nordstrom_et_al_2012,Marijo_Telenta_2014,Hatanaka_et_al_1997,Alhajraf_2004,Wilson_2004,Bouvet_et_al_2006,Santiago_et_al_2007,Benli_Liu_2014,Bitog_et_al_2009} focused on the flow characteristics over a single fence, it has been shown {by means of wind-tunnel experiments \citep{Guan_et_al_2009}, as well as Computational Fluid Dynamic simulations \citep{Lima_et_al_2017}, that large-scale arrays consisting of many fences (5-10 fences or more) exhibit larger wind speeds near the ground between fences far downwind in the array}. Moreover, the simulations by \cite{Lima_et_al_2017}, which calculated the average turbulent wind flow in a two-dimensional cut along the symmetry axis of the fences, showed how maximal wind speeds occurring within the array depend on fence porosity, spacing and height. However, to more {realistically model} scenarios with fences of finite cross-wind width, three-dimensional simulations are required.

In the present work, we {simulate the turbulent flow} over a three-dimensional array of sand fences. Our aim is to investigate how the total area of the soil that is protected against wind erosion depends on the main parameters of the system, i.e. fence porosity, height and spacing. {{In the present work, we focus on the structure of the average turbulent wind field over the terrain in presence of the sand fences, and in the absence of saltation or suspension transport. We note that the presence of sand grains in the transport layer alter the wind profile, as showed, for instance, by \cite{Xu_et_al_2018}. However, our main objective is to investigate how different arrays modify the soil area where the local wind velocity (in the absence of saltating grains) is reduced below the minimal threshold value for saltation transport.}} To define this area, we consider the average wind velocity close to the ground within the region between the fences. We want to provide useful insights e.g.~for developing strategies to optimize the design of an array of fences. One of the challenges in such strategies consists in achieving maximal soil protection with the smallest amount of material for fence construction and maintenance \citep{Lima_et_al_2017}. {In this manuscript, we thus provide comparative results} from simple arrays of fences (i.e. when the fence height and inter-fence spacing are constant over the field) with those from complex arrays of fences with multiple values of height and spacing.

\section{Numerical experiments}

\begin{figure}[htpb]
\centering
\includegraphics[width=0.9\linewidth]{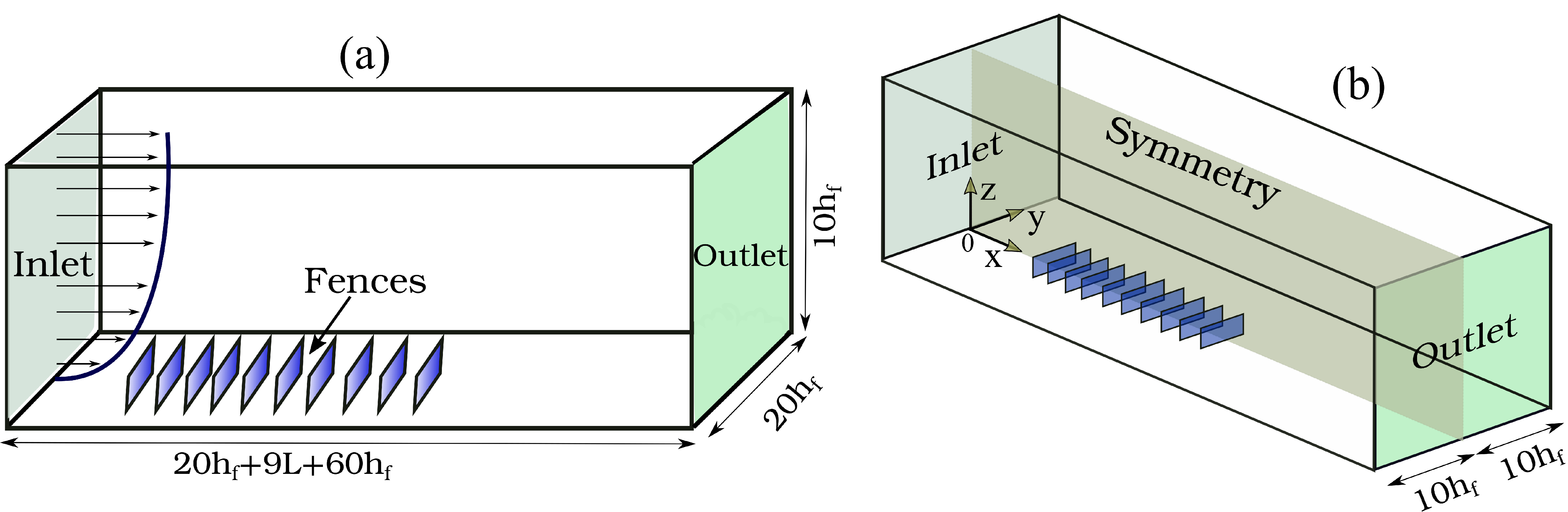}
\caption{Schematic diagram showing the geometry of the wind channel with the array of fences on its bottom. The fences have height $h_{\mathrm{f}}$ and between them a spacing $L$. At the inlet, the logarithmic profile for the wind velocity is imposed following Eq.~(\ref{eq:wind_profile}).}
\label{fig:simulation_setup}
\end{figure}

{Figure \ref{fig:simulation_setup} shows} the schematic representation of the {setup employed in our simulations}. {This consists} of a three-dimensional channel of height ${\Delta}z = 10\,h_{\mathrm{f}}$, width ${\Delta}y=20\,h_{\mathrm{f}}$ and length ${\Delta}x = 80h_{\mathrm{f}}+9L$, while the fences, each of height $h_{\mathrm{f}}$, are erected vertically on the bottom wall of this channel at different values of inter-fence spacing $L$, as described later. {The fences have a cross-wind width $W_{\mathrm{f}} = 10\,h_{\mathrm{f}}$}, which is $50\%$ of the lateral width ${\Delta}y$ of the wind channel (Fig.~\ref{fig:simulation_setup}). The soil level in the absence of {the} fences is constant and equal to zero, while the dimensions of the wind channel are chosen to be large enough to ensure that border effects can be neglected, {but small enough for efficient numerical simulation}. We have checked that the results, presented in the next section, remained unchanged by increasing the box dimensions. {{We use a full size model for the sand fences, i.e. our simulations are performed with heights $h_{\mathrm{f}} = $ $50\,$cm and $1\,$m which are consistent with values applied in field experiments of aeolian soil protection \citep{Li_and_Sherman_2015}.}}

We consider that the fences are subjected to an average turbulent wind flow in $x$ direction (cf.~Fig.~\ref{fig:simulation_setup}). In the absence of fences, the average horizontal wind velocity over the flat ground ${\vec{u}}(x,y,z)$ increases logarithmically with the height $z$ above the bed level \citep{Bagnold_1941,Pye_and_Tsoar_1990}. To represent such average upwind flow condition in our calculations, the logarithmic velocity profile for ${\vec{u}}(x,y,z)$ is imposed as boundary condition at the simulation inlet. Specifically, at the horizontal position $x=0$, ${\vec{u}}(0,y,z) = 0$ for $z = \delta$, where $\delta$ is the surface roughness. The horizontal velocity increases with $z$ according to the following equation (which is valid for $z \geq \delta$) \citep{Bagnold_1941,Pye_and_Tsoar_1990}: 
\begin{equation}
{\vec{u}}(0,y,z) = \frac{u_{{\ast}0}}{\kappa}{\mathrm{log}}{\left[{{\frac{z}{\delta}}}\right]}{\vec{e}}_x, \label{eq:wind_profile}
\end{equation}
where ${\vec{e}}_x$ is the unit vector pointing in the direction $x$, $u_{{\ast}0}$ is the magnitude of the upwind shear velocity of the wind and $\kappa = 0.4$ is the von K\'arm\'an constant. The shear velocity $u_{{\ast}0}$ is proportional to the mean flow velocity gradient in {turbulent boundary layer flow, and} is used to define the upwind shear stress, 
\begin{equation}
{\vec{\tau}}_0 = {\vec{\tau}}(0,y) = \left[{{\rho}_{\mathrm{air}}{u_{{\ast}0}^2}}\right]{\vec{e}}_x, \label{eq:tau}
\end{equation}
where ${\rho}_{\mathrm{air}} = 1.225\,$kg$/$m$^3$ denotes the air density. Moreover, we take for the surface roughness the value $\delta = 100\,{\mu}$m, {which is within the range between $10\,{\mu}$m and $1.0\,$mm valid for dune fields \citep{Pye_and_Tsoar_1990}. } {{In our simulations, we apply a wind shear velocity $u_{{\ast}0} = 0.4\,$m/s, considering that sand-moving upwind shear velocities in dune fields can reach from the threshold for sustained saltation, i.e. $u_{{\ast}{\mathrm{t}}} \approx 0.2\,$m$/$s, up to values of the order of $0.5\,$m$/$s \citep{Sauermann_et_al_2003,Claudin_et_al_2013}. Moreover, much stronger wind velocities of the order of $4\,u_{{\ast}{\mathrm{t}}}$ are not considered since above this value sand grains are transported in suspension \citep{Paehtz_et_al_2013}.}}

In our calculations, each sand fence is modeled as a vertical, porous wall, implemented by a special type of boundary condition representing a porous membrane of a certain velocity/pressure drop characteristics \citep{Wilson_1985,Santiago_et_al_2007,Araujo_et_al_2009,Yeh_et_al_2010}. At height $z$, this pressure drop is given by the expression
\begin{equation}
{\Delta}p(x,y,z) = -\frac{1}{4{\Phi}^2}{\rho}_{\mathrm{air}}[\vec{u}(x,y,z) \cdot {\vec{e}}_x]^2{{\Delta}m}, \label{eq:pressure_drop}
\end{equation}
where the term in the brackets is the wind velocity normal to the fence, i.e.~the horizontal wind speed at height $z$, ${\Delta}m$ is the fence's thickness and $\Phi$ its porosity. The fences' thickness is set as ${\Delta}m = 10^{-4}\,$m, while a nominal porosity of $20\%$ \citep{Li_and_Sherman_2015,Lima_et_al_2017} is used in our calculations. 

{{We remark that our model does not resolve the pores, i.e. it is a continuum model where the pores are assumed to be uniformly distributed throughout the fence cross-sectional area. Accordingly, our model does not address the effect of the shape of the opening (pores). However, \cite{Li_and_Sherman_2015} summarized literature data on the effect of the shape and size of the pores, and showed that the effect of pore shape and size is non-trivial but small, and their impact is most often reflected in the changing characteristics of the turbulence in the immediate lee of a fence. The investigation of the shape of the opening and the arrangement of the pores certainly poses an interesting topic for future modeling.}}

We consider that the fluid (air) is incompressible and Newtonian, while the calculation of the average turbulent wind field over the soil is performed as described in previous works \citep{Herrmann_et_al_2005,Araujo_et_al_2013}. Specifically, the Reynolds-averaged Navier-Stokes equations are solved using the FLUENT Inc.~commercial package (version 14.5.7), in which the standard $\kappa-\epsilon$ model is applied in the computations to simulate turbulence. The time-averaged (or Reynolds-averaged) Navier-Stokes equations for the wind flow over the terrain are solved in the fully-developed turbulent regime. In the calculations, a non-slip boundary condition is applied to the fluid-solid interface defined by the soil and the fences, while at the top wall, the shear stress of the wind is set equal to zero \citep{Herrmann_et_al_2005,Almeida_et_al_2006,Almeida_et_al_2008,Araujo_et_al_2013,Michelsen_et_al_2015,Lima_et_al_2017}. Since the fences have a finite cross-wind width $W_{\mathrm{f}} = 0.5 \cdot {\Delta}y$ (see Fig.~\ref{fig:simulation_setup}), to avoid border effects and to resolve the flow in the region around the edges of the fences, both edges of the fence array are separated from the lateral walls by a distance equal to half the fence width (see Fig.~\ref{fig:simulation_setup}). Given the symmetry of the system, and for the sake of computational efficiency, a symmetric boundary condition is applied in the $y$ direction, with symmetry plane $y = 0$. The flow equations are, thus, solved for the right half of the simulation domain, and the developed solution mirrored along the symmetry plane $y = 0$ to obtain the flow field on the left half. Both lateral walls are treated as full slip walls, i.e.~having zero shear stress. 

Moreover, the default pressure-velocity coupling scheme (``SIMPLE'') of the solver, {which obtains a correction in the static pressure field such as to satisfy the continuity equation \citep{Patankar_and_Spalding_1972}}, is applied with its preselected values of parameters. {At the outlet, a static pressure $P=0$ is specified}, while the default option ``standard wall functions'' of the solver is selected \citep{Araujo_et_al_2013,Lima_et_al_2017}. This option applies the wall boundary conditions to all variables of the $k-\epsilon$ model that are consistent with Eq.~(\ref{eq:wind_profile}) along the channel's bottom wall \citep{Launder_and_Spalding_1974}. 

To perform the calculations, a second-order upwind discretization scheme is applied to the momentum, whereas for the turbulent kinetic energy and turbulence dissipation rate we apply a first-order upwind scheme \citep{Araujo_et_al_2013}. A rectangular grid with mean spacing of about $0.05\,{\mathrm{m}}$ is used for the lower region from the bottom wall ($z=0$) up to the height of the fences ($z=h_{\mathrm{f}}$), while for larger heights, a coarser grid is considered. Specifically, the grid cell size is $0.10\,{\mathrm{m}}$ within the range $h_{\mathrm{f}} \leq z < 2h_{\mathrm{f}}$ and $0.50\,{\mathrm{m}}$ for $2h_{\mathrm{f}} < z \leq 10h_{\mathrm{f}}$ (i.e. up to the top wall). {{Analysis prior to our previous studies \citep{Araujo_et_al_2013,Lima_et_al_2017} showed little changes in the flow profile by decreasing cell size below the aforementioned values, while too large cell sizes affected the convergence of the solution. However, we remark that an interesting point for future work is the investigation of how different meshes affect surface roughness and flow convergence. Such an investigation could be conducted, for instance, by explicitly accounting for mesh elements at the inlet, such as in previous wind tunnel experiments \citep{Xu_et_al_2018}.}}

Moreover, the following initial conditions are applied: for all values of $x,y,z$, the velocity and the pressure are set to zero, while at the left wall ($x=0$), the logarithmic profile Eq.~(\ref{eq:wind_profile}) is imposed. Convergence of the numerical solution of the transport equations is evaluated in terms of residuals, which provide a measure for the degree up to which the conservation equations are satisfied throughout the flow field. Here, convergence is achieved when the normalized residuals for both $\epsilon$ and $k$ fall below $10^{-4}$, and when the normalized residuals for all three velocity components (in the directions $x$, $y$ and $z$) fall below $10^{-6}$.

\section{Results and discussion}

\begin{figure}[htpb]
\centering
\includegraphics[width=0.73\linewidth]{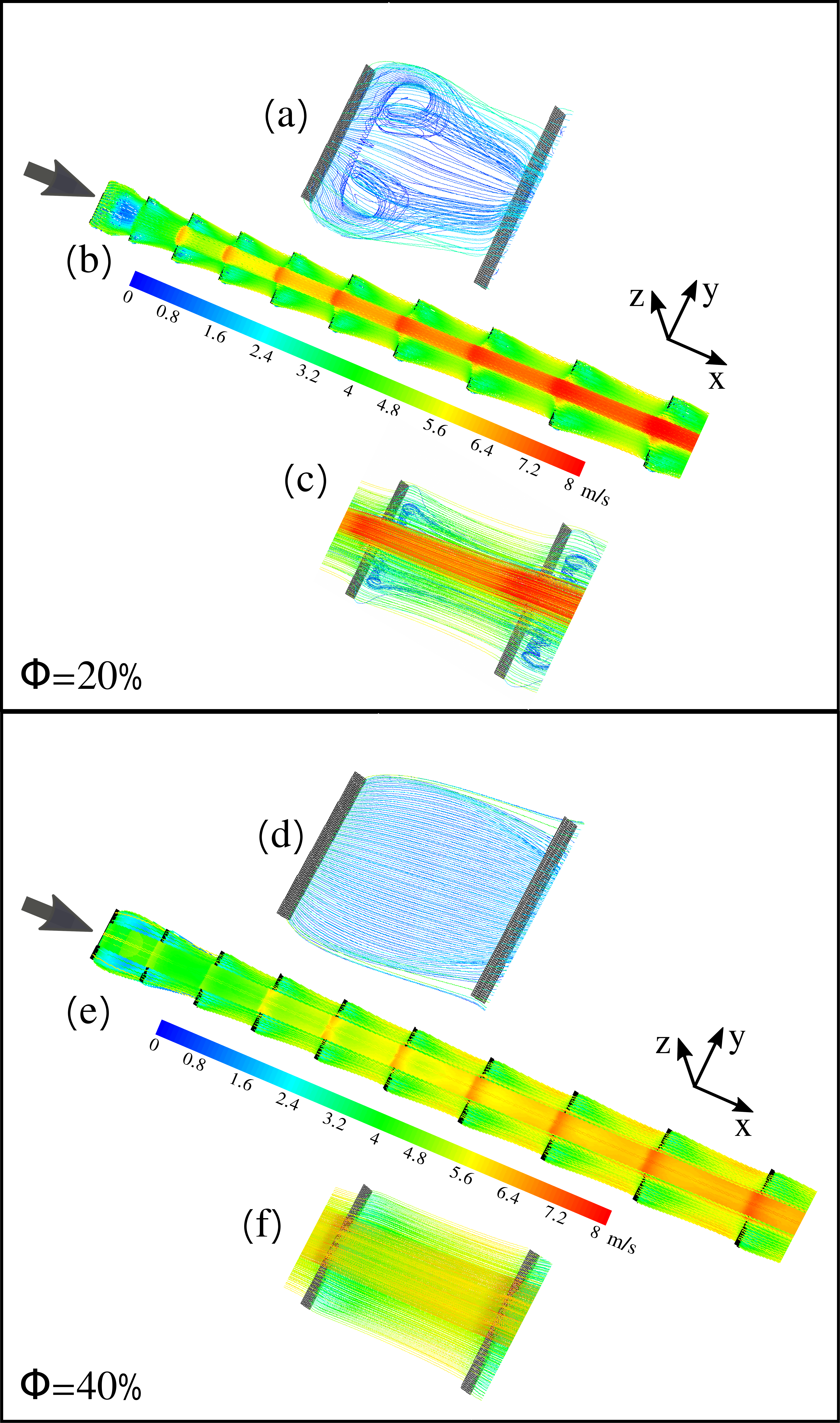}
\caption{Streamlines of the average turbulent wind flow over an array of 10 fences, obtained for {{porosity $\phi = 20\%$ (top) and $\phi = 40\%$ (bottom). The streamlines over the entire array obtained with these two porosity values are shown in (b) and (e), respectively. In (a) and (d) we see the respective streamlines for the first pair of fences, while (c) and (f) denote the corresponding results for the last two fences. The arrow at left in each panel indicates the wind direction. The colors indicate the magnitude of the wind velocity in m/s.}} {{Upwind shear velocity of the wind is $u_{{\ast}0} = 0.4\,$m$/$s}} and fence height is $50\,$cm, while fence spacing is $5\,$m. }
\label{fig:streamlines}
\end{figure}

Fig.~\ref{fig:streamlines} shows {flow streamlines} over an array of 10 fences of height $h_{\mathrm{f}}=0.5\,$m, spacing $L_x = 20\,h_{\mathrm{f}} = 10\,$m and {{two values of porosity: $\phi = 20\%$ (upper panel in Fig.~\ref{fig:streamlines}), and $\phi = 40\%$ (lower panel in Fig.~\ref{fig:streamlines})}}. In our simulations, the cross-wind width of the fences is $W_{\mathrm{f}}=5\,$m. The wind velocity upwind of the fences, $u_{{\ast}0}$, is $0.4\,$m$/$s, and its direction is indicated by the arrow in Fig.~\ref{fig:streamlines}. The streamlines shown in this figure clearly indicate the increase in the internal boundary layer due to the presence of the fences, which has been discussed previously from two-dimensional CFD simulations considering the two-dimensional cut along the symmetry line of the fences \citep{Lima_et_al_2017} (see also \cite{Guan_et_al_2009,Jerolmack_et_al_2012}). Here, in the three-dimensional simulation shown in Fig.~\ref{fig:streamlines} {{for the porosity ${\phi} = 20\%$ (Fig.~\ref{fig:streamlines}a-c), we see the occurrence of flow recirculation  on the horizontal plane, which is not visible in the simulation results for $\phi = 40\%$ (Fig.~\ref{fig:streamlines}d-f). This is consistent with experimental and field observations of enhanced turbulence and occurrence of whirls for very low values of porosity \citep{Li_and_Sherman_2015}. The porosity $\phi = 40\%$ is within the range of fence porosities that have proven to be most efficient in reducing flow pressure and turbulent kinetic energy over the sheltered terrain, i.e. $30\% < \phi < 50\%$ \citep{Lee_and_Lim_2001,Park_and_Lee_2003,Li_and_Sherman_2015}.}}

\subsection{Longitudinal profile of the wind shear velocity near the surface} 

\begin{figure}[htpb]
\centering
\includegraphics[width=0.7\linewidth]{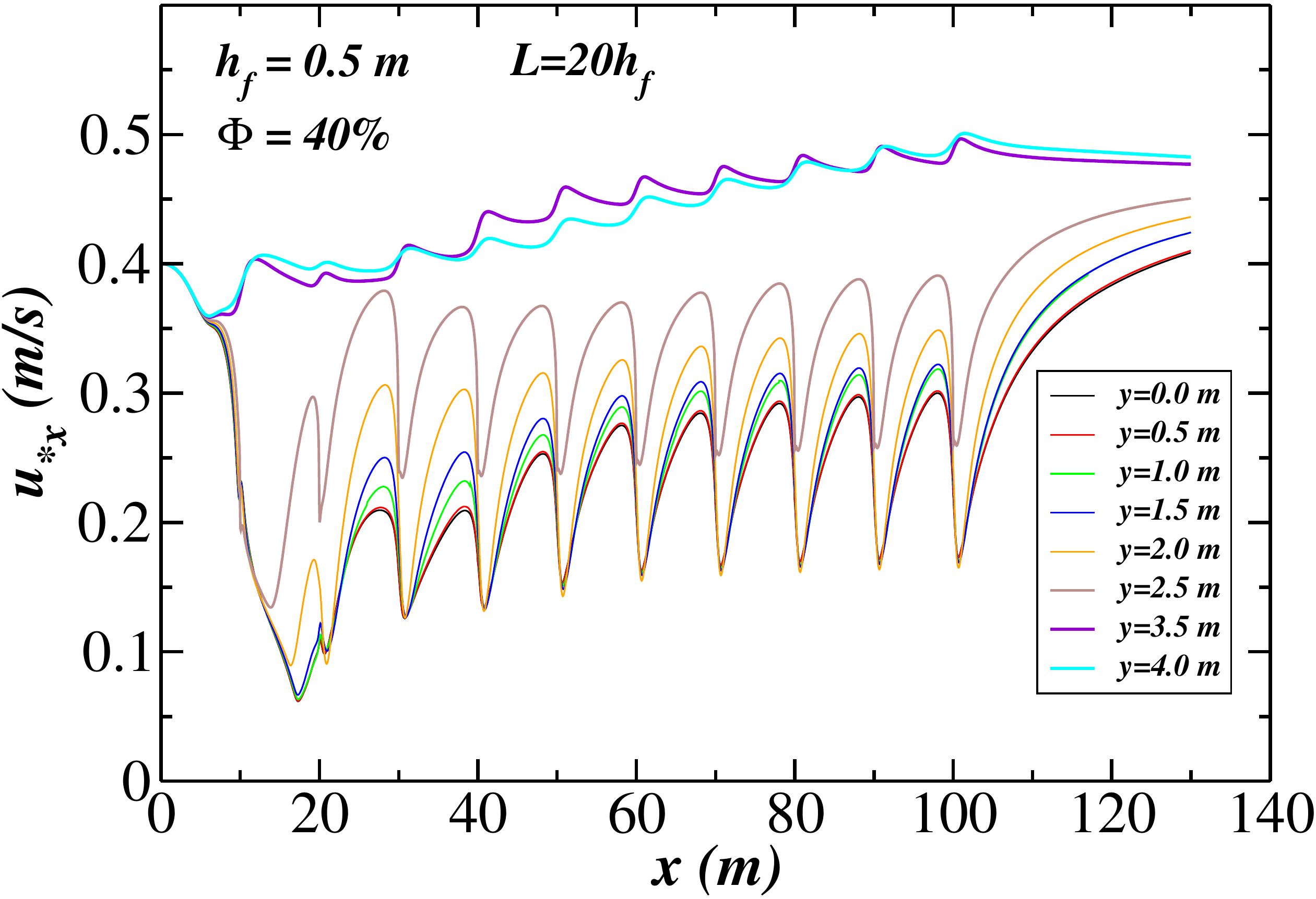}
\caption{Average longitudinal component of the wind shear velocity, $u_{{\ast}x}$, calculated using Eq.~(\ref{eq:u_star_x_definition}), as a function of the downwind position $x$ for different values of $y$. The first fence is at $x = 10\,$m, the last one at $x=100\,$m and the spacing is $L_x=10\,$m. Fence porosity is $40\%$, fence cross-wind width is $5\,$m and fence height is $50\,$cm, while the upwind shear velocity is $u_{{\ast}0}=0.4\,$m$/$s. }
\label{fig:u_star_x}
\end{figure}

We calculate, for different longitudinal slices of the system, i.e.~slices in the direction of the wind, the wind shear velocity $u_{{\ast}x}(x,y)$ as a function of the downwind position $x$ close to the ground. {{\cite{Guan_et_al_2009} have shown, by means of two-dimensional simulations, that, at the reference height of $0.2\,h_{\mathrm{f}}$, the reduction in the horizontal wind velocity $u_x(x)/u_x(x=0)$ as a function of the downwind position $x$ provides a reasonable approximation for the factor $u_{{\ast}x}(x)/u_{{\ast}x}(x=0)$, i.e.~$u_{{\ast}x}(x)/u_{{\ast}0}$, as long as the soil properties are the same for the surface between the fences and upwind the fence array. Therefore,}} following the experimental work by \cite{Guan_et_al_2009}, we use the relation 
\begin{equation}
u_{{\ast}x}(x,y) \approx u_{{\ast}0}r(x,y), \label{eq:u_star_x_definition}
\end{equation}
where the two-dimensional field {$r(x,y)$ is defined as $r(x,y) \equiv u_x(x,y,z=0.2\,h_{\mathrm{f}})/u_x(0,y,z=0.2\,h_{\mathrm{f}})$, with $u_x(x,y,z) = \vec{u}(x,y,z) \cdot \vec{e}_x$ denoting the longitudinal component of the wind velocity. $r(x,y)$ gives thus the attenutation in the longitudinal wind velocity due to the presence of the fences, computed at a height of $z = 0.2\,h_{\mathrm{f}}$ above the soil at the position $(x,y)$.} By doing so, we obtain the results shown in Fig.~\ref{fig:u_star_x}, which correspond to the same parameter values of the simulation in Fig.~\ref{fig:streamlines}.

We see in Fig.~\ref{fig:u_star_x} that the wind shear velocity in the region between the fences near the symmetry axis of the system ($y = 0$) is reduced by a substantial amount. In particular, between the last two fences in the system, the maximal value of $u_{{\ast}x}$ at $y = 0$ is about $63\%$ of the upwind shear velocity $u_{{\ast}0}$, while the wind shear velocity increases with the lateral position towards the flanks of the fences. The two uppermost curves in Fig.~\ref{fig:u_star_x} correspond to values of $|y| > 2.5\,$m, that is beyond the lateral borders of the fences. We see that in this outer region, the wind shear velocity is largest thereby exceeding the upwind value $u_{{\ast}0}$ after the third fence. This result can be understood by noting that the fence array imposes an obstacle for the wind, and that there is thus a pressure reduction at the lateral border of the fence array. This pressure reduction is associated with the conservation of momentum of the fluid flow in the system, which leads to an increase in wind velocity near the flanks of the fence array. 

\begin{figure}[htpb]
\centering
\includegraphics[width=0.7\linewidth]{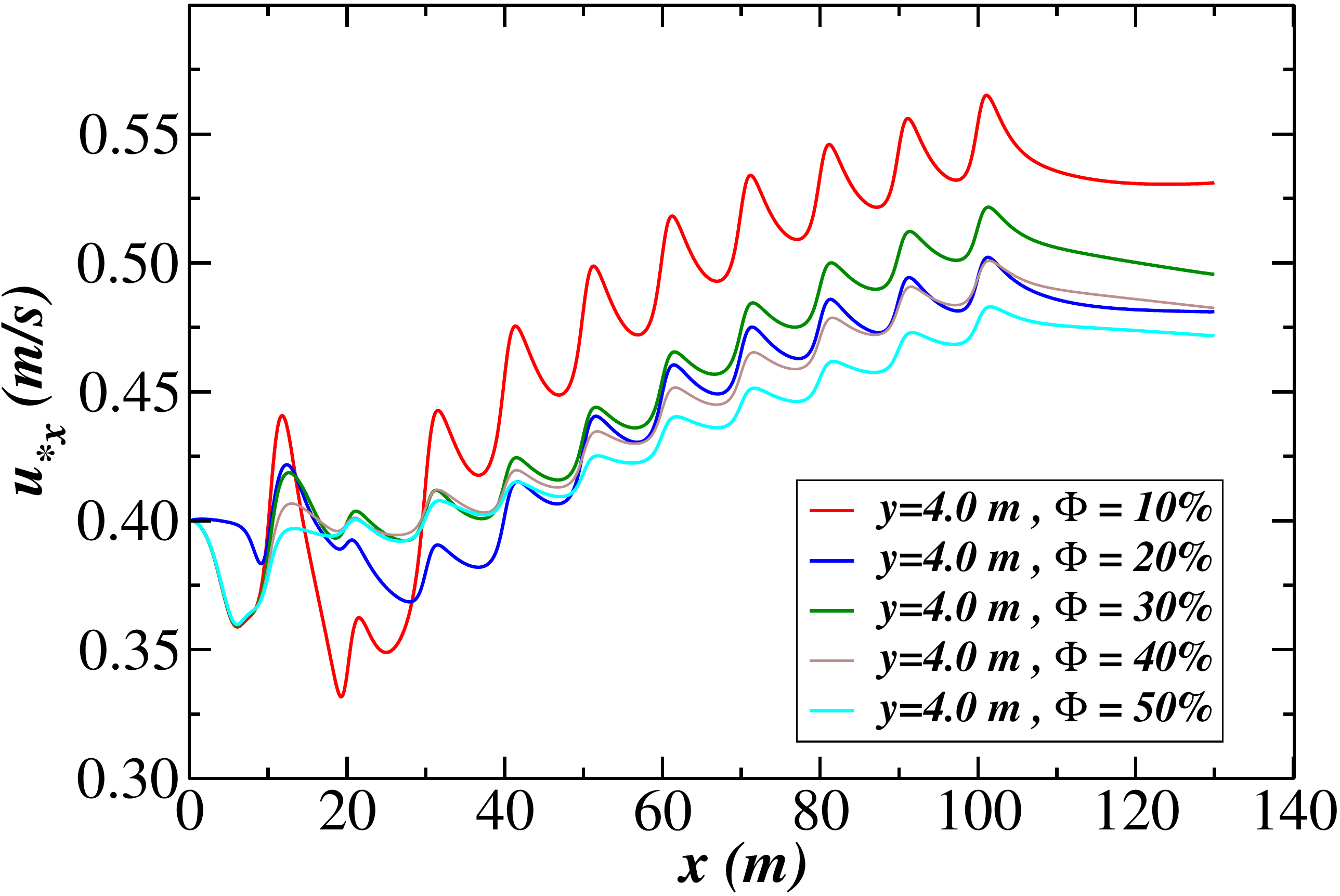}
\caption{Average longitudinal component of the wind shear velocity, $u_{{\ast}x}$, calculated using Eq.~(\ref{eq:u_star_x_definition}), as a function of the downwind position $x$ for $y=4\,$m and different values of porosity $\phi$. Fence positions and sizes are as in Fig.~\ref{fig:u_star_x}. Moreover, the upwind shear velocity is $u_{{\ast}0}=0.4\,$m$/$s.}
\label{fig:flow_lateral_border}
\end{figure}

Moreover, we see in Fig.~\ref{fig:flow_lateral_border} that the behavior of $u_{{\ast}x}$ along the flanks of the fence array depends on the fence porosity $\phi$. The increase in flow velocity at the lateral borders is stronger the higher the porosity, since the pressure drop at the sides of the fence array is stronger the less permeable the fences. {{We note that the objective of the analysis in Fig.~\ref{fig:flow_lateral_border} is to investigate the effect of the porosity on the local wind shear velocity near the flanks of the fences. To this end, we evaluate in Fig.~\ref{fig:flow_lateral_border} the behavior of $u_{{\ast}x}$ along the longitudinal slice $y = 4\,$m, which has been chosen here since it is still close to the fence flanks but far enough from the lateral border of the channel.}}

\subsection{Area protected against sand transport}

One important aspect of a sand fence array is its efficiency to protect soil against motion of sand grains. Since the impact of sand particles on the ground during saltation causes dust emission \citep{Shao_et_al_1993}, soil protection against sand transport has implications not only for the dynamics of sand encroachment and aeolian desertification, but also for the Earth's climate and atmosphere.

\begin{figure}[htpb]
\centering
\includegraphics[width=1.0\linewidth]{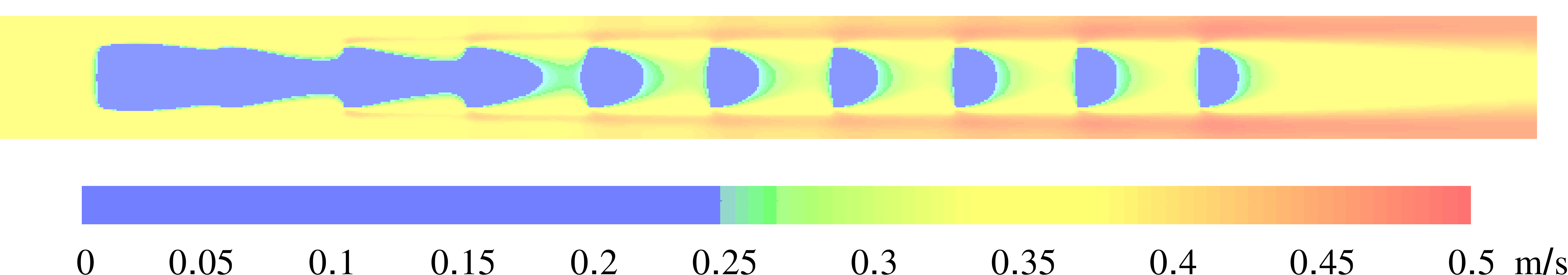}
\caption{Longitudinal component of the wind shear velocity field, $u_{{\ast}x}$, calculated using Eq.~(\ref{eq:u_star_x_definition}), for the array of fences in Fig.~\ref{fig:u_star_x}. The colors indicate the values of $u_{{\ast}x}$ in m/s.}
\label{fig:soil_protection}
\end{figure}

Fig.~\ref{fig:soil_protection} shows the two-dimensional field of the longitudinal component of the wind shear velocity close to ground as a function of the horizontal position, $u_{{\ast}x}(x,y)$. The blue area in the figure represents the wake region within the fence array, i.e.~values of $u_{{\ast}x}(x,y)$ smaller than $0.25\,$m$/$s, which is approximately the minimal threshold shear velocity for direct entrainment of sand particles by fluid forces \citep{Bagnold_1941}. We see in Fig.~\ref{fig:soil_protection} that the shape of the wake region changes with distance downwind. For the first upwind fences, wake zones produced by the adjacent fences appear connected to each other, but they disconnect after the fourth fence.  

\begin{figure}[htpb]
\centering
\includegraphics[width=0.5\linewidth]{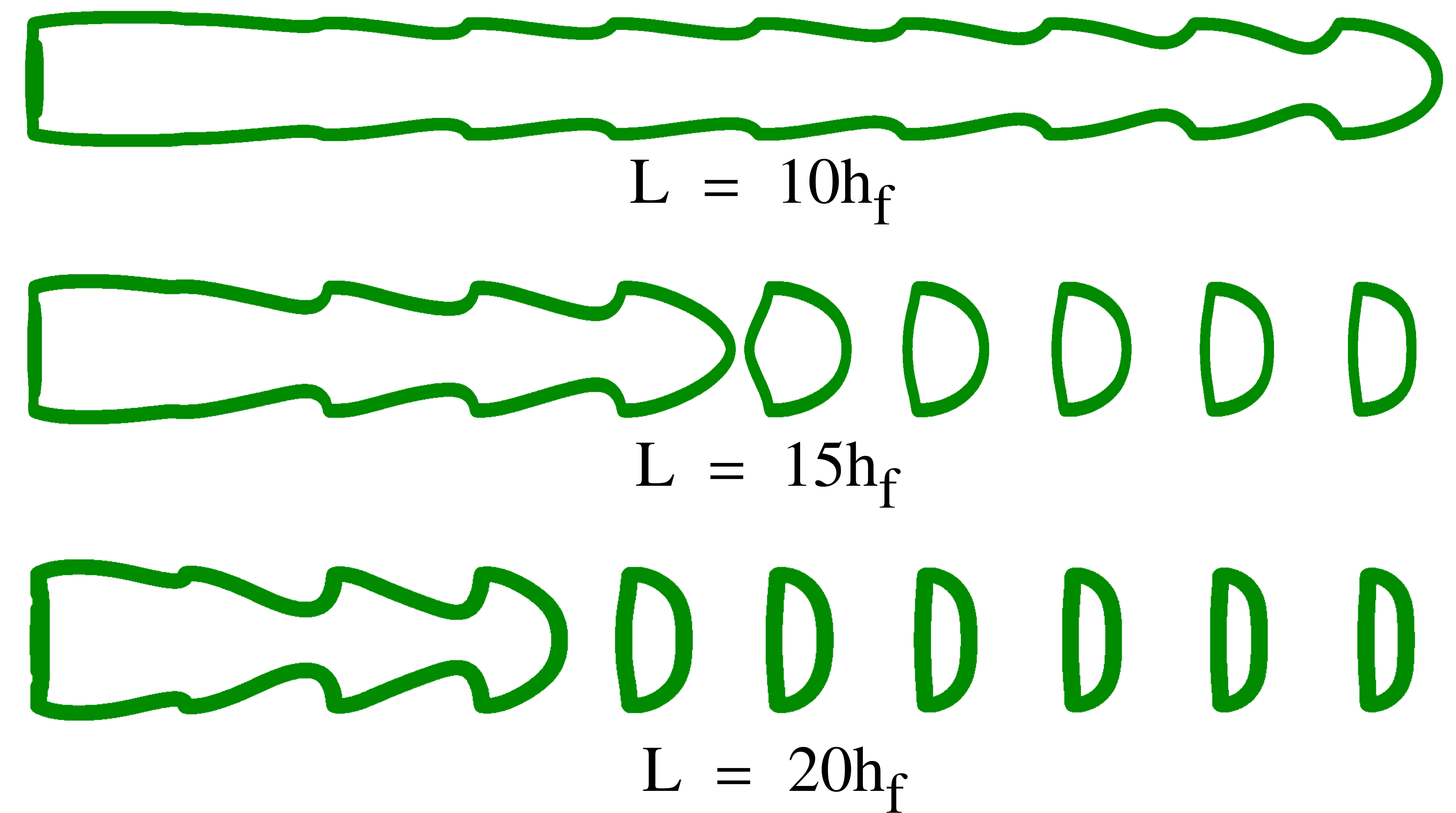}
\caption{Contour plot of the protected soil area, i.e. the area within which $u_{{\ast}x} < u_{\mathrm{ft}}$ ($=0.25\,$m$/$s), for different spacings between the fences. The other parameters are the same as in Fig.~\ref{fig:u_star_x}.}
\label{fig:contour_soil_protection}
\end{figure}

Moreover, the shape of the wake zones as a function of the downwind distance, as well as the downwind position where the wake zones separate from each other,  depend on several parameters of the system, such as upwind flow velocity $u_{{\ast}0}$, inter-fence distance $L_x$, fence porosity $\phi$ and number of fences in the array. Figure \ref{fig:contour_soil_protection} shows how the wake zones depend on the spacing between the fences. This figure shows that, for $u_{{\ast}0}=0.4\,$m$/$s, {{$\phi = 40\%$}} and 10 fences, separation of the different wake zones of the fences starts after a certain value of spacing which is within the range $15 < L_x/h_{\mathrm{f}} < 20$. 

\begin{figure}[htpb]
\centering
\includegraphics[width=0.48\linewidth]{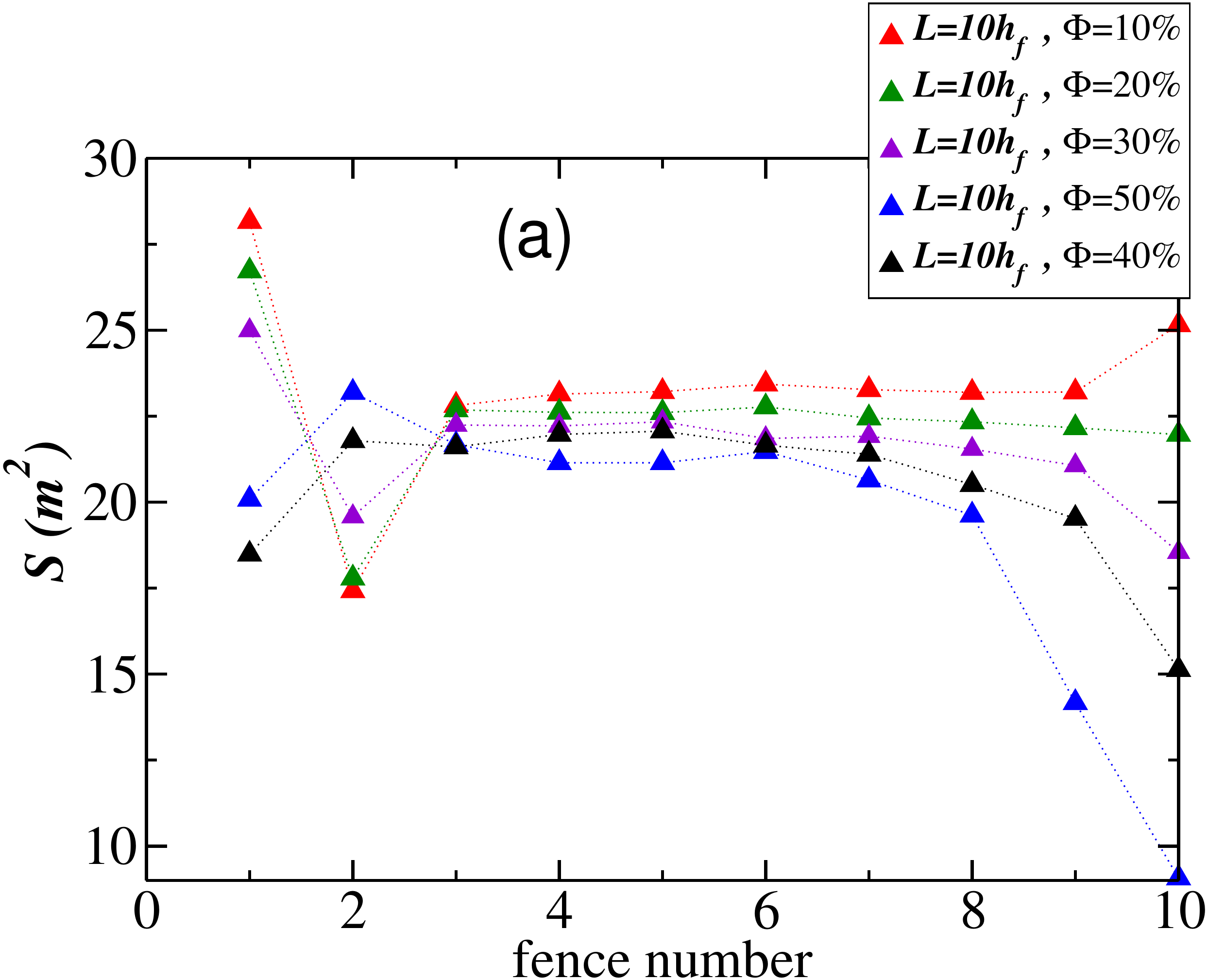}
\includegraphics[width=0.48\linewidth]{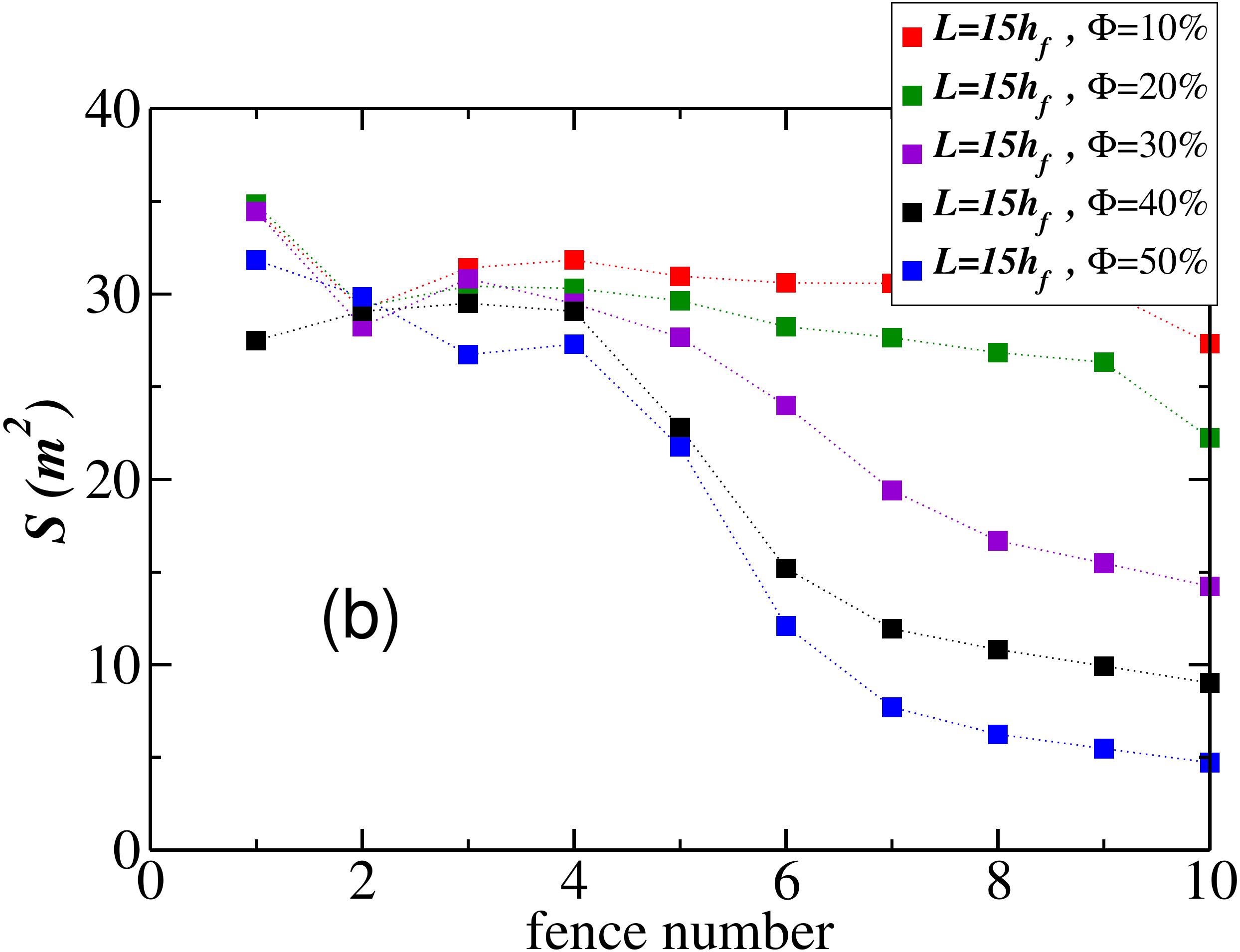}
\includegraphics[width=0.48\linewidth]{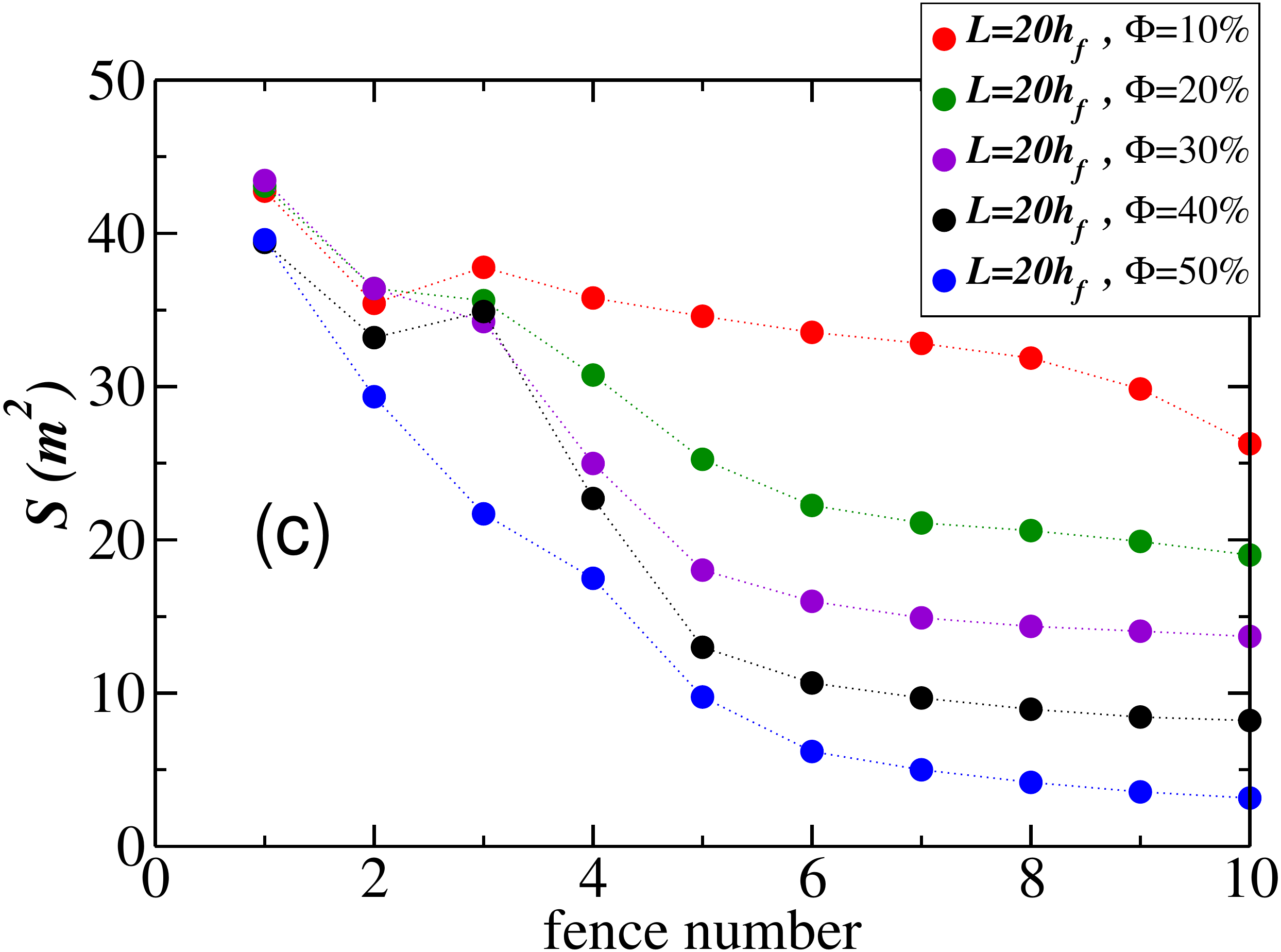}
\caption{Protected soil area $S$ between two adjacent fences, i.e. the area within which $u_{{\ast}x} < u_{\mathrm{ft}}$ ($=0.25\,$m$/$s), as a function of the fence position downwind for different values of the spacing and porosity. The area $S$ is defined according to Eq.~(\ref{eq:S}) and is rescaled by $S_0 = {\Delta}_y \cdot L$, which is the area between two adjacent fences.}
\label{fig:S}
\end{figure}

We investigate this behavior in more detail by varying the fence porosity and spacing (see Figs.~\ref{fig:S}a-c). The fence height is constant and equal to $50\,$cm, while the upwind flow shear velocity is $0.4\,$m$/$s \citep{Lima_et_al_2017}. Each one of Figs.~\ref{fig:S}a-c shows the total protected area $S$, for which $u_{{\ast}x} < u_{\mathrm{ft}}$ ($=0.25\,$m$/$s), as a function of the fence number $\#i$. This area $S$ is measured from the mid positions between the two neighbouring fences, that is,
\begin{equation}
S = \int_{-0.5\,{\Delta}_y}^{0.5\,{\Delta}_y}\int_{x_i-0.5\,L_x}^{x_i+0.5\,L_x} {\Theta}(u_{{\ast}x}(x,y)-u_{\mathrm{ft}}) \cdot dx\,dy \label{eq:S}
\end{equation}
where $\Theta(x)$ is the Heaviside function, i.e. $\Theta(x) = 0$ for $x \leq 0$ and unity otherwise, while $\Delta_y$ is the cross-wind width of the channel and $L_x$ the fence spacing.

For a spacing of $10\,h_{\mathrm{f}}$ (Fig.~\ref{fig:S}a), we see that, for all porosity values from $10\%$ to $50\%$, the wake region consists of interconnected wake zones forming a protected soil region of approximately constant shape and size throughout the fence array. This behavior leads to an approximately constant value of $S$ associated with each fence, although for very high porosities the flow through the last fences tends to recover upwind conditions earlier, thus leading to a drop in the value of $S$. By taking a larger value for the fence spacing (cf.~Fig.~\ref{fig:S}b; $L_x = 15\,h_{\mathrm{f}}$), the fence wake zones appear inter-connected in the beginning of the array (up to the fourth fence). We see that, in this initial region, the value of $S$ is approximately the same for all values of porosity (see Fig.~\ref{fig:S}b). However, from the fourth fence onwards, the zones of shear velocity below threshold become smaller due to the obstacle's permeability, and $S$ starts to decrease downwind at a rate that increases with fence porosity. This behavior is associated with a thinning of the wake region with distance downwind as can be seen in Fig.~\ref{fig:contour_soil_protection}).

Moreover, Fig.~\ref{fig:S}b shows that, for high enough porosities, $S$ first decreases rapidly with distance downwind (between fences 4 and 6) and then slowly from the sixth fence onwards, thus indicating the approach to an asymptotic value for very large downwind distances. This asymptotic behavior corresponds to the regime of disconnected wake zones, which is achieved faster the larger the fence spacing (see Fig.~\ref{fig:contour_soil_protection}). Furthermore, we calculate in Fig.~\ref{fig:S}c the value of $S$ as a function of the fence position for a higher spacing of $20\,h_{\mathrm{f}}$. Clearly, the spacing in these calculations is so large that a dependence of $S$ on the porosity is observed from the very beginning of the array, i.e. $S$ decreases faster downwind the larger $\phi$. The separation of the wake zones produced by the fences occurs from the sixth fence onwards for all values of $\phi$ larger than $10\%$. Far downwind within the array, the system is found within the regime (or phase) of separated / disconnected wake zones (regime B), which is separated from the upwind regime of inter-connected wake zones (regime A) at some point within the array depending on the inter-fence spacing.

\subsection{Two-fences experiments elucidate the critical dependence of wake zones connectivity on inter-fence spacing}

To investigate the dependence of the wake zones between adjacent fences on the spacing, described above, we perform systematic calculations of the protected soil area between an isolated pair of fences for different values of spacing. To this end, we perform a different type of numerical experiments in which only 2 fences are considered, while the spacing between both fences is systematically varied. The two fences have the same height, porosity and cross-wind width as the fences in the array considered in the previous calculations. Moreover, the wind tunnel has the same height and width ${\Delta}y$, and boundary conditions are the same as before. The only difference compared to the setup of Fig.~\ref{fig:simulation_setup} is the channel's downwind length, which is now $80h_{\mathrm{f}}+L$.

\begin{figure}[htpb]
\centering
\includegraphics[width=0.5\linewidth]{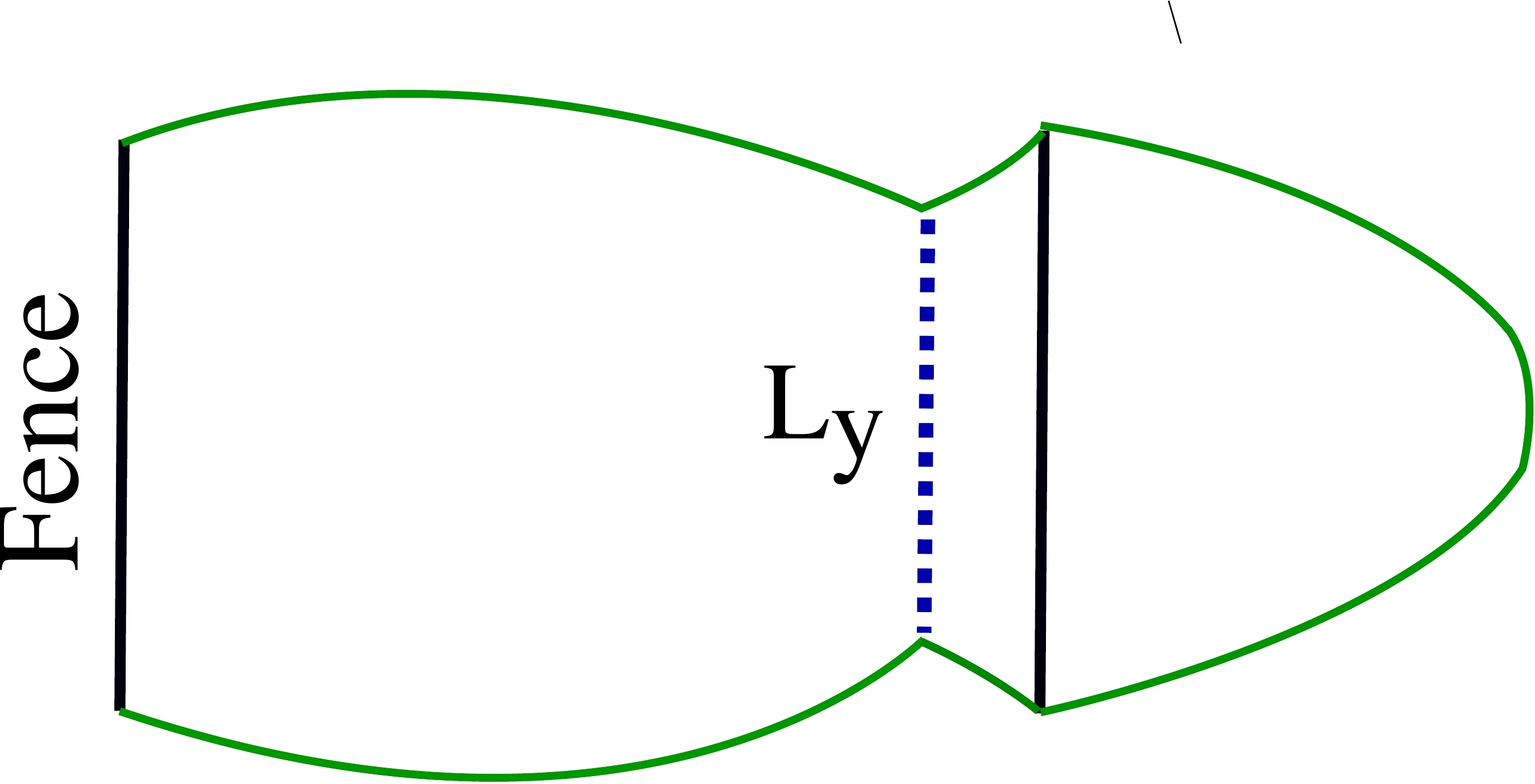}
\caption{Contour of the area $S$ protected against aeolian entrainment in the two-fences experiment, obtained for $u_{{\ast}0} = 0.4\,$m$/$s, $h_{\mathrm{f}}=50\,$cm, $\phi=40\%$ and $L_x=20\,h_{\mathrm{f}}$ (spacing). $L_y$ denotes the smallest cross-wind width of the protected region between the fences.}
\label{fig:2_fences}
\end{figure}

Figure \ref{fig:2_fences} shows the contour plot of the wake zone from the upwind fence including the entire protected area associated with the downwind fence. In this particular calculation, the porosity is $40\%$, the fence height is $50\,$cm, the spacing is $10\,h_{\mathrm{f}}$ and the upwind shear velocity is $0.4\,$m$/$s. 

\begin{figure}[htpb]
\centering
\includegraphics[width=0.7\linewidth]{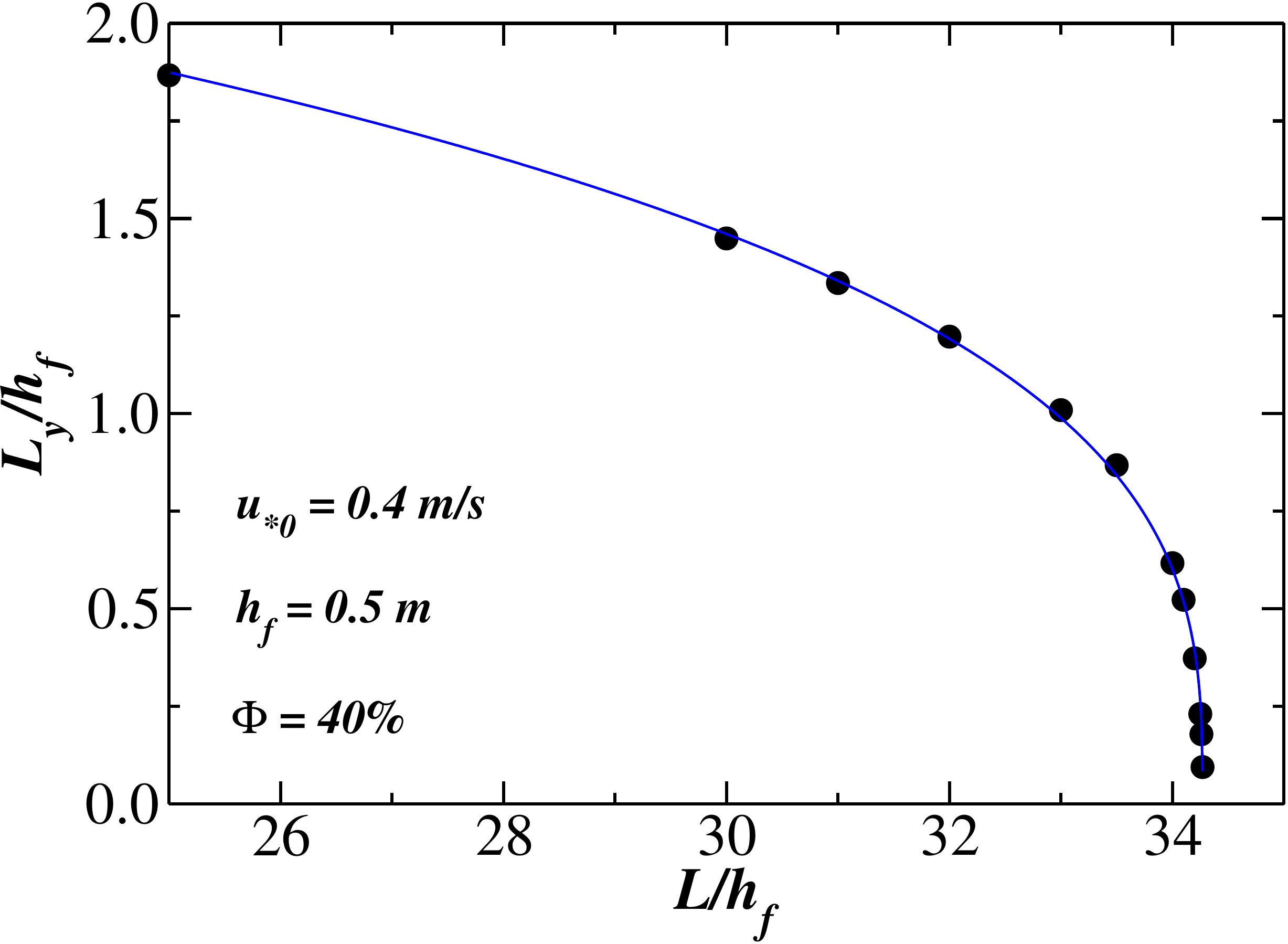}
\caption{Value of $L_y$ defined in Fig.~\ref{fig:2_fences} as a function of the spacing between the fences, $L_x$, in the two-fences experiment. Both $L_y$ and $L_x$ are normalized by the fence height $h_{\mathrm{f}}$. The filled circles denote our calculation data. The continuous line represents the best fit using Eq.~(\ref{eq:Ly}), which gives ${\tilde{L}}_{xc} \approx 34.3$, $A \approx 2.85$ and $\beta \approx 0.32$.}
\label{fig:Ly_Lx}
\end{figure}

In Fig.~\ref{fig:2_fences}, $L_y$ denotes the smallest cross-wind width of the wake zone between the two fences. This cross-wind width decreases with the fence spacing $L_x$ as shown in Fig.~\ref{fig:Ly_Lx}. We see from this figure that there is a critical value of $L_x/h_{\mathrm{f}}$, which is slightly larger than $34$, beyond which $L_y$ is zero, i.e. the wake zones of the two fences are separated from each other. For larger values of the spacing ($\tilde{L}_x \equiv L_x/h_{\mathrm{f}} \gtrsim 34$), the system is in the disconnected phase (regime B; $\tilde{L}_y \equiv L_y/h_{\mathrm{f}} = 0$), while as the spacing decreases ($\tilde{L}_x \lesssim 34$), the system enters regime A, i.e. the connected phase ($\tilde{L}_y \neq 0$). 

This behavior is similar, for instance, to the transition of percolation from the connected to the disconnected phase as the probability $p$ that a site in the graph is open decreases below a critical value $p_c$ \citep{Herrmann_and_Roux_1990,Aharony_and_Stauffer_1994,Bunde_and_Havlin_1996,Araujo_et_al_2002,Parteli_et_al_2010}. In other words, in the system of Fig.~\ref{fig:Ly_Lx}, the cross-wind width of the wake zone between the fences, $L_y$, goes to zero with a power law at $\tilde{L}_x = \tilde{L}_{xc}$ around $34$ from regime B to regime A. We can thus describe the behavior of ${\tilde{L}}_y$ by means of the following equation,

\begin{align}
{\tilde{L}}_y &= 
\begin{cases}
A \cdot {\left[{1 - {\tilde{L}}_x/{{{\tilde{L}}_{xc}}}}\right]}^{\beta}, & {\tilde{L}}_x < {\tilde{L}}_{xc} \\
0, & {\tilde{L}}_x \geq {\tilde{L}}_{xc} \label{eq:Ly}
\end{cases} 
\end{align} 
where the critical point ${\tilde{L}}_{xc}$, the constant $A$ and the exponent $\beta$ can be determined from the best fit to the simulation data. The continuous line in Fig.~\ref{fig:Ly_Lx} shows this best fit, which gives ${\tilde{L}}_{xc} \approx 34.3$, $A \approx 2.85$ and $\beta \approx 0.32$. The quality of the fit is remarkable. We thus conclude that the change in the system's behavior from regime A to regime B exhibits the characteristics of a second order phase transition at the critical spacing ${\tilde{L}}_{xc}$, where ${\tilde{L}}_y$ represents an appropriate observable (order parameter) to describe this transition. 

{The results of our simulations are important for the drag partitioning schemes in models for aeolian surfaces with sand fences. In regime B, i.e.~$\tilde{L}_x > {\tilde{L}}_{xc}$, the total drag can be then partitioned into a pressure drag, due to the momentum absorbed by the individual fences, and the surface drag on the underlying surface \citep{Raupach_1992}. However, in regime A, i.e.~$\tilde{L}_x < {\tilde{L}}_{xc}$, interactions of the turbulent wakes and mutual sheltering among the fences (see Fig.~\ref{fig:contour_soil_protection}) lead to a reduction in the pressure drag on individual fences. In other words, as the density of the roughness elements {{(in the present study, the sand fences)}} increases, individual elements become less effective since only a fraction of them can be seen by the mean flow \citep{Shao_and_Yang_2005,Yang_and_Shao_2006}. As shown by \cite{Shao_and_Yang_2005}, the effect of mutual sheltering between the roughness elements can be taken into account by including in the drag partitioning scheme by \cite{Raupach_1992} a third component, which is the skin drag due to momentum transfer to the surfaces of roughness elements \citep{Shao_and_Yang_2005,Yang_and_Shao_2006}.}  

{However,} it is interesting to discuss the disconnection of the wake zones between two adjacent fences within an array (cf.~Fig.~\ref{fig:contour_soil_protection}) in the light of the analysis made for two isolated fences in Figs.~\ref{fig:2_fences} and \ref{fig:Ly_Lx}. {{In other words, these figures concern simulations of two fences, so the following question arises: How the results presented in these figures relate to multiple fence simulations?}} For a fence array, we have found that the wind shear velocity is increasing downwind, which means that ${\tilde{L}}_y$ decreases with distance downwind. The larger the spacing of the array, the earlier within the array the transition from regime A to regime B (disconnected wake zones) will occur, which means that the wind shear velocity is one important parameter controlling the critical spacing ${\tilde{L}}_{xc}$. 

{{Therefore, considering, for instance, the array of 10 fences, qualitatively, the same behavior of the protected area as a function of inter-fence spacing observed in Figs.~\ref{fig:2_fences} and \ref{fig:Ly_Lx} occurs for any pair of fences in the array, but the critical spacing for the transition between regimes A and B decreases with distance downwind. Moreover, in a given array, this critical spacing depends on several factors including the fence porosity and height, incident wind velocity and number of fences, since the disturbance of the boundary layer depends on the array's length and number of roughness elements (the fences). This dependence is, thus, worth investating in more detail in future work. Moreover, the present study concerns a wind of constant velocity inciding perpendicularly on the fences, and thus further investigation is required to address realistic scenarios of wind velocity and trend variations.}} However, the analysis presented here represents a first step towards a complete description of the complex behavior of the protected soil area as a function of the geometric properties of an array of sand fences.

\subsection{\label{sec:multiple_heights}Multiple or constant fence heights?}

\begin{figure}[htpb]
\centering
\includegraphics[width=1.0\linewidth]{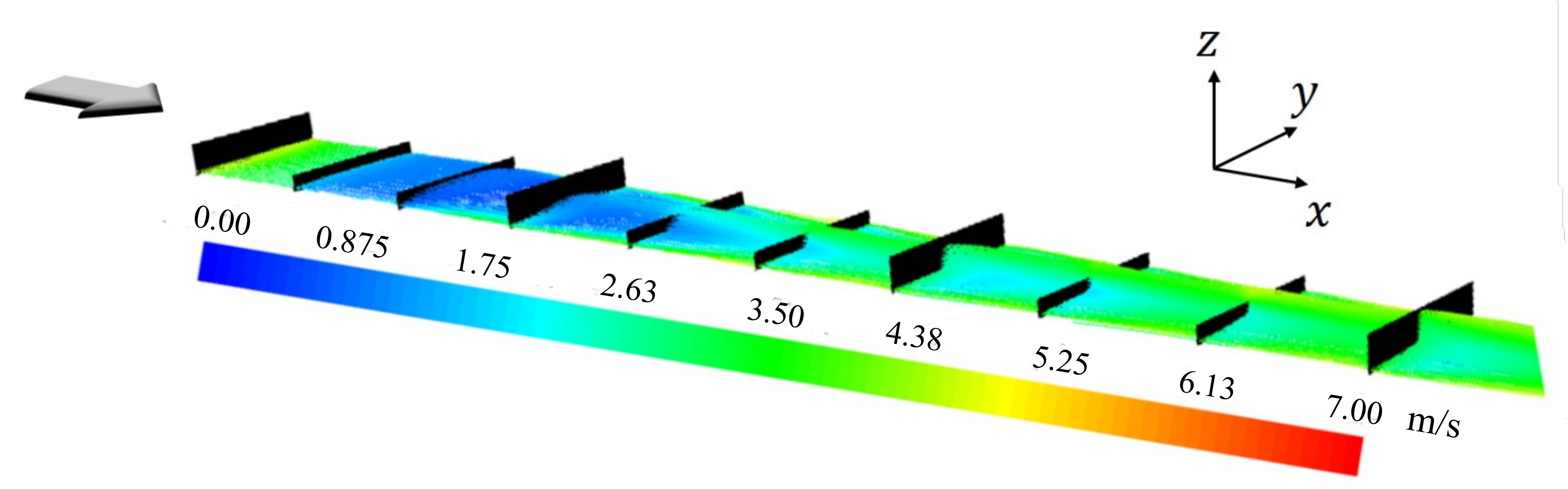}
\caption{Flow streamlines obtained for setup B, i.e. fences of height $1\,$m separated by pairs of fences of height $50\,$cm with constant spacing $6.67\,$m. The colors indicate the magnitude of the wind velocity in m/s. Upwind flow velocity $u_{{\ast}0} = 0.4\,$m$/$s and fence porosity $\phi = 40\%$.}
\label{fig:multiple_fences_streamlines}
\end{figure}

In this subsection, we address the question whether combining fences of different heights may increase the efficiency of the fence array. Using the same amount of fence material, the aim is to find a combination of fence heights for which soil protection against aeolian erosion is optimal. To characterize soil protection, we consider the total amount of protected area ($S$) and we analyze different configurations of fences using two different heights, namely $h_{\mathrm{f}} = 50\,$cm and $h_{\mathrm{f}}=1\,$m, defined as follows:
\begin{itemize}
\item setup A --- fences of height $h_{\mathrm{f1}} = 1.0\,$m separated by a distance $10h_{\mathrm{f1}}$ ($= 10\,$m)
\item setup B --- fences of height $h_{\mathrm{f1}} = 1.0\,$m separated by pairs of fences of height $h_{\mathrm{f2}} = 50\,$cm, with constant spacing $6.67\,$m between all fences (see Fig.~\ref{fig:multiple_fences_streamlines}) 
\item setup C --- fences of height $h_{\mathrm{f1}} = 1.0\,$m separated by pairs of fences of height $h_{\mathrm{f2}} = 50\,$cm, with spacing determined by the height of the upwind fence --- in this setup, each pair of adjacent fences within the array is separated by a distance equal to 10 times the height of the upwind fence
\item setup D --- fences of height $h_{\mathrm{f2}} = 50\,$cm separated by a distance $10h_{\mathrm{f2}}$ ($= 5\,$m)
\end{itemize}

{{We note that the inter-fence spacing of $6.67\,$m in setup B was chosen to conserve the same amount of fence material (fence cross-sectional area) for the sake of comparison between all setups. Therefore, using setup B with the referred spacing leads to the same amount of fence material as in the other setups, for the same total target array's along-wind size.}}

\begin{figure}[htpb]
\centering
\includegraphics[width=1.0\linewidth]{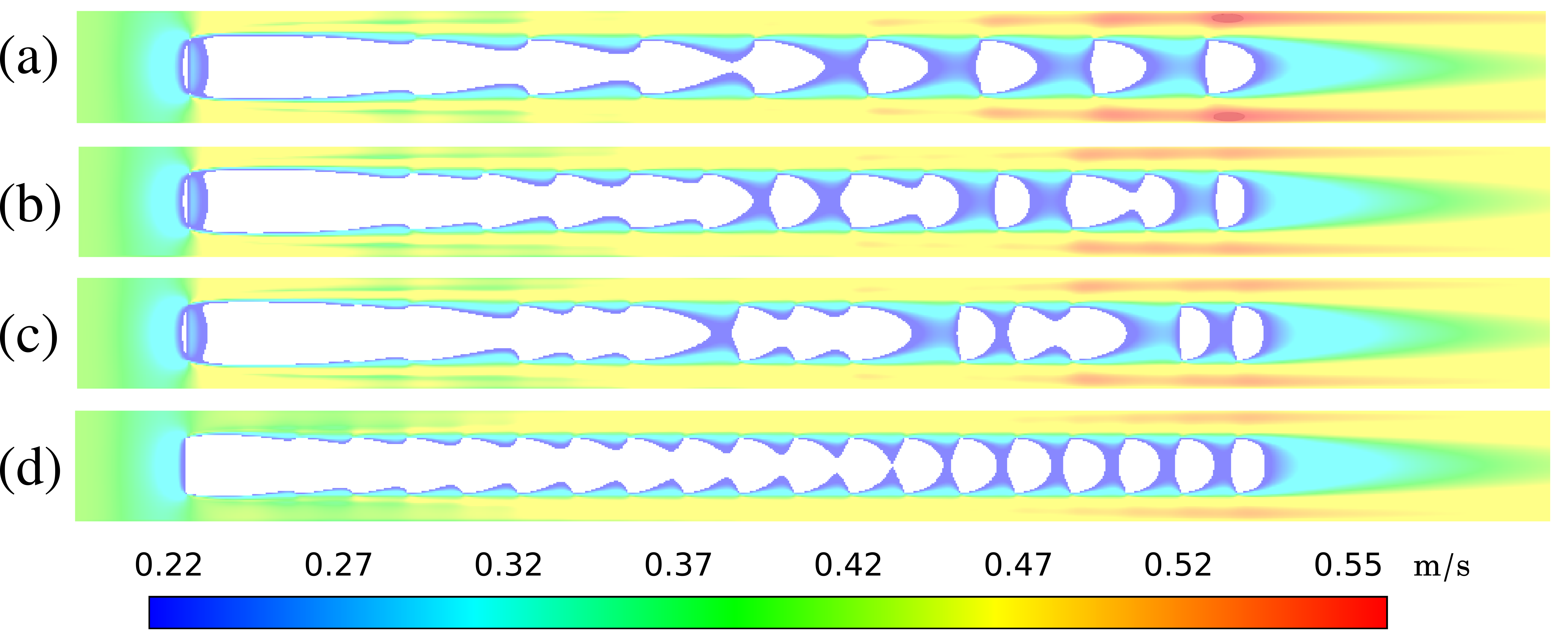}
\caption{Longitudinal component of the wind shear velocity field, $u_{{\ast}x}$, for the arrays of fences of setups A-D using multiple fence heights. {{{\em{Setup A}}: $h_{\mathrm{f}} = 1.0\,$m; {\em{Setup B}}: Alternating heights, homogeneous; {\em{Setup C}}: Alternating heights, heterogeneous; {\em{Setup D}}: $h_{\mathrm{f}} = 50\,$cm.}} The colors indicate the values of $u_{{\ast}x}$ in m/s. The calculation results correspond to upwind flow velocity $u_{{\ast}0} = 0.4\,$m$/$s and fence porosity $\phi = 40\%$, while fence height and inter-fence spacing are varied over the different setups as described in Section \ref{sec:multiple_heights}.}
\label{fig:velocity_field_multiple_heights}
\end{figure}

\begin{figure}[htpb]
\centering
\includegraphics[width=0.9\linewidth]{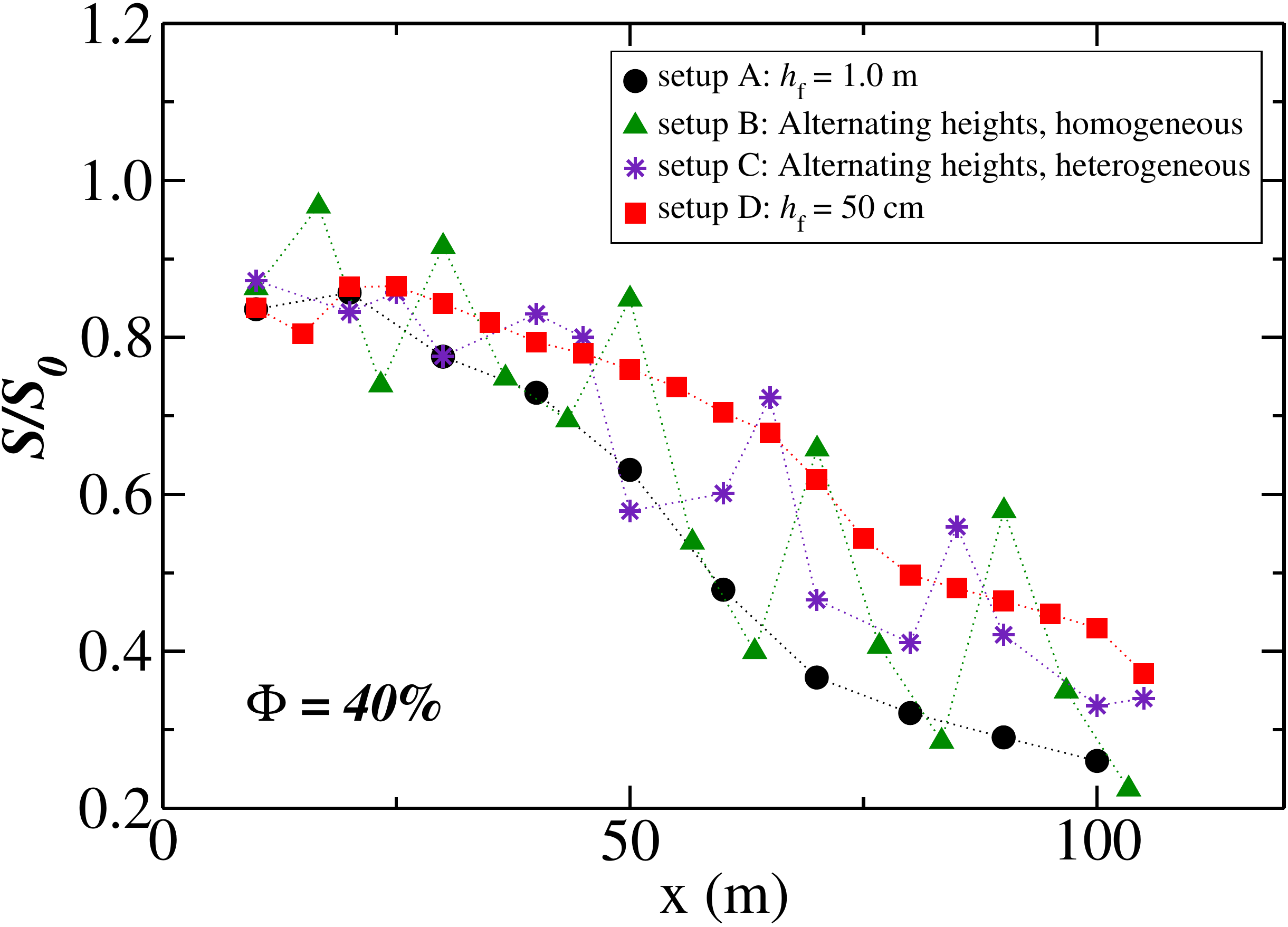}
\caption{Area $S$ of soil between adjacent fences protected against aeolian entrainment as a function of the downwind position, computed for the different setups A-D using multiple fence heights. The points in each curve correspond to the fences $i=1...N$ in the respective setups A-D, while $S$ is the protected area between fence $i$ and $i+1$, i.e. the area between the adjacent fences within which $u_{{\ast}x} < u_{{\ast}{\mathrm{ft}}}$. Moreover, $S_0$ represents the total area ${\Delta}_y \cdot L_i$ between fences $i$ and $i+1$.}
\label{fig:S_multiple_heights}
\end{figure}

Figure \ref{fig:velocity_field_multiple_heights} shows the two-dimensional velocity field of the horizontal wind shear velocity $u_{{\ast}x}(x,y)$ for setups A-D, by using a fence porosity of $40\%$ and upwind shear velocity $0.4\,$m$/$s. The total area $S$ for which $u_{{\ast}x} < u_{\mathrm{ft}}$ is calculated, as described in the previous subsection, as a function of the downwind position for all setups, and the result is shown in Fig.~\ref{fig:S_multiple_heights}. In this figure, the protected area produced by each fence is shown rescaled by the total area $S_0$ between the fence and its downwind neighbour. We can see that setup D, which uses constant fence height of $50\,$cm, produces larger values of $S/S_0$ compared to setup A (constant fence height of $1\,$m). Moreover, this figure shows that setups using multiple fence heights lead to intermediate efficiency between the two investigated setups of constant fence height. Nevertheless, we see that, at some positions along the array, setups B and C locally produce lower $S/S_0$ than setup A and higher $S/S_0$ than D, and the efficiency of the array should be thus analysed by considering an average over the entire surface on which the fences have been erected.

We see in Fig.~\ref{fig:velocity_field_multiple_heights} that the unprotected soil zones in the different setups ($u_{{\ast}x} > u_{\mathrm{ft}}$) display very distinct behaviour with respect to the distribution of wind shear velocities. In particular, the ``gaps'' between the wake zones in setup A are larger thus giving rise to higher transport rates between the fences in setup A than in setup D. We thus expect that using setup A may lead to larger transport rates compared, for instance, to setup D. To verify this behavior, we calculate the following quantity for the different setups,
\begin{equation}
Q = \int_{-0.5\,{\Delta}_y}^{0.5\,{\Delta}_y}\int_{10}^{100} {\Theta}(u_{{\ast}x}(x,y)-u_{\mathrm{ft}}) \cdot [u_{\ast}^2(x,y) - u_{{\mathrm{ft}}}^2]dx\,dy \label{eq:Q}
\end{equation}
which determines, up to a pre-factor encoding parameters related to sediment and fluid properties, the saturated sediment flux with the wind shear velocity \citep{Lettau_and_Lettau_1978,Kok_et_al_2012}. More precisely, we calculate the ratio $Q/Q_0$ for all setups, where $Q_0$ is given by Eq.~(\ref{eq:Q}) with $u_{\ast}(x,y) = u_{{\ast}0}$, that is the undisturbed wind shear velocity in the absence of the dunes. We note that the sand flux in Eq.~(\ref{eq:Q}) incorporates important simplifications, in particular neglecting the hystheresis in sediment transport --- i.e. the correct threshold wind velocity in the flux equation would be the impact threshold, which is about $80\%$ of $u_{\mathrm{ft}}$ \citep{Bagnold_1941,Kok_et_al_2012}. Moreover, Eq.~(\ref{eq:Q}) also neglects the effect of the saturation length of sediment transport on the downwind relaxation of the flux toward its saturated value \citep{Sauermann_et_al_2001,Kroy_et_al_2002}. Nevertheless, the present analysis is useful to investigate the trend in the transport rate with respect to the fence array setups defined above.

Our calculations yield the results given in Table \ref{tab:Q_Q0}. Clearly, setup D performs best with respect to sand flux reduction. 
\begin{table}[ht]
\centering
\caption{\label{tab:Q_Q0} Sand flux reduction due to sand fences arranged according to the different setups. The sand flux in the presence of the fences has been calculated using Eq.~(\ref{eq:Q}), while $Q_0$ denotes the sand flux without the presence of the fences ($u_{{\ast}x} = u_{{\ast}0} = 0.4\,$m$/$s throughout the field).}
\begin{tabular}{|l||c|c|c|c|}
\hline
setup & A & B & C & D \\
\hline
$Q/Q_0$ & $0.120$ & $0.074$ & $0.083$ & $0.052$ \\
\hline
\end{tabular}
\end{table}

We have also investigated the average wind shear velocity within the disturbed boundary layer in the region above the fences, i.e.~at height between $3\,h_{\mathrm{f}}$ and $8\,h_{\mathrm{f}}$ \citep{Guan_et_al_2009}. Within this region, wind shear velocities are larger than $u_{{\ast}0}$ throughout the array of fences, with the speedup decreasing downwind towards an asymptotic value \citep{Guan_et_al_2009,Jerolmack_et_al_2012}. However, to the best of our knowledge, our study is the first systematic investigation of the two-dimensional field $u_x(x,y)$ associated with the near-surface wind shear velocity on the protected soil. We hope that the present study will motivate future investigation on the effect of different system parameters on the protected area $S$, as well as on the critical rescaled distance ${\tilde{L}}_{xc}$, the constant $A$ and the exponent $\beta$ in Eq.~(\ref{eq:Ly}). Moreover, it would be interesting to verify our prediction for the phase transition in Eq.~(\ref{eq:Ly}) by means of field measurements or wind tunnel experiments.

\section{Conclusions}

We performed three-dimensional CFD simulations of the average turbulent wind field over a flat terrain in presence of an array of porous fences. The calculations were performed with different values for porosity, height and spacing, and the present work focused on the total area of sediment soil protected against aeolian entrainment. We found that there is a critical inter-fence spacing $L_{xc}$ that separates two regimes of behavior: For $L_x > L_{xc}$, the wake zones of the different fences in the array are disconnected (regime B), while for $L_x < L_{xc}$, no separation occurs between zones of wind shear velocity below threshold (regime A). We find that the system undergoes a continuous phase transition from regime B to regime A as $L_x$ decreases below $L_{xc}$. In regime A, the cross-wind width of the protected zone, $L_y$ (the order parameter) scales with $(1-L_x/L_{xc})^{\beta}$, where the exponent $\beta$ is around $0.32$. 

{{Moreover, we have analyzed the influence of fence's height and porosity on the fluid structure. Our simulations have shown that an array of 50\,cm high fences performs better, with regard to minimization of soil erosion and sand flux, than counterparts containing 1\,m high fences or fence heights alternating between 50\,cm and 1\,m. Furthermore, we have found that, along the flanks of the fence array, the along-wind component of the wind shear velocity $u_{{\ast}x}$ is largest and increases linearly with distance downwind. In particular, $u_{{\ast}x}$ exceeds the upwind value $u_{{\ast}0}$ already after the third fence. Our simulations have shown that the increase in flow velocity at the lateral borders is stronger the higher the porosity, since the pressure drop at the sides of the fence array is stronger the less permeable the fences.}} 

{{The techniques we investigated in the present manuscript have impact on man-made geomorphology. Specifically, the investigation performed in our work is concerned with an intervention that is designed to prevent soil erosion, i.e. the erection of sand fences to reduce the local wind speed below the minimal value for sediment transport.}} One outlook of the present work would be the simulation of the soil topography in presence of the fences. Such simulations can be achieved by coupling the CFD modeling presented here with the morphodynamic model for aoelian landscapes developed in the last years \citep{Sauermann_et_al_2001,Kroy_et_al_2002,Sauermann_et_al_2003,Parteli_et_al_2006,Herrmann_et_al_2008,Duran_et_al_2010,Parteli_et_al_2011,Luna_et_al_2009,Luna_et_al_2011,Luna_et_al_2012,Melo_et_al_2012,Parteli_et_al_2014_EPJST,Parteli_et_al_2019_EGU,Muchowski_et_al_2019_GSGS}. 

Moreover, many open questions remain to be investigated in the future, such as how the predictions reported here change if intermittent transport conditions that occur in real field scenarios \citep{Ellis_et_al_2012,Sherman_et_al_2013} are considered. Moreover, one further aspect to be incorporated in future studies is the effect of the stochastic nature of turbulence on the threshold for sand and dust emission. Accurately modeling such forces is a difficult task that is still matter of intense research \citep{Klose_and_Shao_2012}, and is important to correctly represent threshold wind speeds for sediment entrainment \citep{Kok_et_al_2012,Li_et_al_2014}. Large-scale arrays of sand fences disposed in different ways, such as in zig-zag \citep{Bitog_et_al_2009} or checkerboards \citep{Qiu_et_al_2004}, should be also modeled in future works. 

{{In this context, we remark that \cite{Xu_et_al_2018} pioneered the study of the flow characteristics of turbulent aeolian sand over straw checkerboard barriers and the erosion-deposition patterns resulting from the interaction between particles, fluid and obstacles. In particular, these authors studied the flow over checkerboards by taking into account the motion of the particles, which disturb the wind profile \citep{Xu_et_al_2018}. In their simulations, the checkerboards consisted of 10\,cm high obstacles, and wind tunnel experiments were also performed using the same conditions as in the simulations to verify the predictions for wind velocity and sand flux \citep{Xu_et_al_2018}. The model by \cite{Xu_et_al_2018} is, thus, different from ours, since we investigate the structure of the flow without particles. It is, thus, not possible to apply the same technique proposed by \cite{Xu_et_al_2018} to validate the simulation predictions for vertical sand flux profile, since in our simulations we only investigated the structure of the turbulent wind free of saltating or suspended particles. Moreover, with regard to a verification of the wind profile alone (without particles), to the best of our knowledge, there is no experiment that investigates the turbulent flow profile using the same conditions as in our simulations. However, the longitudinal profile of the wind shear velocity we have presented in our manuscript matches very well the trend observed in the experiments by \cite{Guan_et_al_2009}. We remark that the study of the flow without particles is important as a preliminary investigation to the sand flux, since the turbulent wind flow has still many complex aspects that are still poorly understood --- one of these aspects is the average shape of the zones with no aerodynamic entrainment as a function of the inter-fence distance, investigated in our manuscript.}}

We hope that the predictions made from the simulations presented here will trigger field work to verify our results, as well as further effort toward optimization of strategies to protect soil erosion with sand fences.

\section*{Acknowledgements}
This work was supported in part by CAPES, CNPq and FUNCAP (Brazilian agencies), the Brazilian Institute INCT-SC, by the German Research Foundation (DFG) Grant RI 2497/7-1 and by ERC Advanced grant FP7-319968 FlowCCS of the European Research Council.

\section*{References}

\bibliography{sample}

\end{document}